\documentclass{article}
\usepackage{graphicx}  
\usepackage{amsmath}   
\usepackage[compress]{cite}
\usepackage{amssymb}   
\usepackage{bm} 
\usepackage{dcolumn}
\usepackage{color}
\usepackage{mathrsfs}
\usepackage{amsfonts}
\usepackage{varioref}
\RequirePackage[colorlinks,citecolor=blue,urlcolor=magenta,linkcolor=blue]{hyperref}
\def\EH{Einstein-Hilbert }
\def\LL{Lanczos-Lovelock }
\def\BY{Brown-York }
\def\gr{general relativity}
\def\RN{Reissner-Nordstr\"{o}m }
\def\KN{Kerr-Newmann }
\labelformat{section}{Section #1} 
\labelformat{subsection}{Section #1} 
\labelformat{subsubsection}{Section #1}
\labelformat{subsubsubsection}{Section #1}
\labelformat{equation}{Eq.~(#1)} 
\labelformat{figure}{Fig.~#1} 
\labelformat{subfigure}{Fig.~\thefigure#1} 
\labelformat{table}{Tab.~#1} 
\labelformat{appendix}{Appendix #1}
\title{Brown-York quasilocal energy in Lanczos-Lovelock gravity and black hole horizons}
\author{Sumanta Chakraborty
\footnote{sumanta@iucaa.in;~~sumantac.physics@gmail.com}\\
{\small{IUCAA, Post Bag 4, Ganeshkhind, Pune University Campus, Pune 411 007, India}}\\
and\\
Naresh Dadhich
\footnote{nkd@iucaa.ernet.in}\\
{\small{Center for Theoretical Physics, Jamia Millia Islamia, New Delhi 110025, India}}\\
{\small{and}}\\
\small{IUCAA, Post Bag 4, Ganeshkhind, Pune University Campus, Pune 411 007, India}}
\begin{document} 
\maketitle

\begin{abstract}
A standard candidate for quasilocal energy in general relativity is the Brown-York energy, which is essentially a two dimensional surface integral of the extrinsic curvature on the two-boundary of a spacelike hypersurface referenced to flat spacetime. Several years back one of us had conjectured that the black hole horizon is defined by equipartition of gravitational and non-gravitational energy. By employing the above definition of quasilocal Brown-York energy, we have verified the equipartition conjecture for static charged and charged axi-symmetric black holes in general relativity. We have further generalized the Brown-York formalism to all orders in Lanczos-Lovelock theories of gravity and have verified the conjecture for pure Lovelock charged black hole in all even $d=2m+2$ dimensions, where $m$ is the degree of Lovelock action. It turns out that the equipartition conjecture works only for pure Lovelock, and not for Einstein-Lovelock black holes.
\end{abstract}
\section{Introduction}\label{BY_Intro}

Defining quasilocal energy for \gr\ is an extremely important but a long eluding problem. Initial attempts in this direction involved pseudotensor methods, leading to coordinate dependent expressions, identification of certain symmetries and defining the Noether charge or some mathematical constructs from the Cauchy data showing similar physical properties associated with energy. Most of these definitions are quite useful and have  physical implications in one case or the other, but no unified description has emerged (for a partial set of references see \cite{Komar:1958wp,Hawking:1968qt,Penrose:1982wp,Dougan:1991zz,Ludvigsen:1981gf,Kulkarni:1988CQG,Bergqvist:1992CQG}). 

The main reason for this ambiguity in defining an energy for gravitational field is due to its non-linear nature and the fact that gravitational energy is non-localizable. This in turn implies not only matter energy produces gravitational field but the gravitational field energy also does the same. A natural definition of quasilocal gravitational energy follows from employing the Hamilton-Jacobi theory for gravitation. This leads to the \BY quasilocal energy as it was first obtained by Brown and York in  \cite{Brown:1992br}. This definition of energy contains, as it should,  contributions from both the matter part and the gravity part. It is defined covariantly and more importantly the energy is additive. Also in the asymptotic limit the \BY energy reproduces the Arnowitt-Deser-Misner mass, providing much credence to  it. This quasilocal energy has been studied extensively in \gr\ for static spacetime with spherical symmetry and also for Kerr black hole, which is stationary 
\cite{Dadhich:1997ku,Dadhich:1997ze,Bose:1998uu,Bose:1999er}. 

One key aspect for \gr\ is that, it simply follows from geometric properties of the Riemann curvature tensor, in particular the Bianchi identity plays a crucial role. Moreover the field equations for \gr\ is second order in the dynamical variable, which ensures that ghost modes do not appear. Then it is interesting to ask whether this setup can be generalized to higher dimensions as well as to higher curvature theories. Remarkably the answer is yes. If we demand that field equations have to be second order in the dynamical variable, the gravitational Lagrangian uniquely picks out the \LL action \cite{gravitation,Padmanabhan:2013xyr,Dadhich:2012ma,pons:2014oya,Dadhich:2015nua}. On the other hand, generalizing the curvature tensor such that trace of its Bianchi derivative vanishes giving the divergence free second rank tensor, which agrees with the one obtained by variation of the action.  This also uniquely leads to the \LL Lagrangian \cite{Dadhich:2008df}. The pure \LL Lagrangian is also intimately connected with spacetime dimensions. For example, field equations for \gr\ is non-trivial for $D>2$ and has free propagation only in $D>3$. For $D=3$, gravity is kinematic, as Riemann tensor is determined entirely by Ricci tensor. Then requiring the kinematic property of gravity to hold in all odd dimensions singles out pure \LL gravity, i.e., one particular order out of the full \LL Lagrangian \cite{Dadhich:2015lra,Dadhich:2012cv,Dadhich:2012pd,Yale:2010jy}. Moreover, from thermodynamic perspectives as well \LL gravity has unique features \cite{Padmanabhan:2009vy}. Most of the thermodynamic results which hold for null surfaces in \gr\ can be generalized to hold in \LL gravity as well \cite{Padmanabhan:2013nxa,Chakraborty:2014rga,Chakraborty:2014joa,Chakraborty:2015wma,Chakraborty:2015hna}. Thus even from the thermodynamic perspective, \LL gravity has a special status. As argued earlier, the quasilocal energy is an important measure of gravitational field, it would therefore be pertinent to study the \BY quasilocal 
energy for \LL gravity as well.

In general, location of black hole horizon is defined as a limit for timelike world lines to exist (when $4$-velocity turns null) and when a spatial surface turns one way membrane (i.e., it can be crossed in one direction only). It has been argued in \cite{Dadhich:1997ze} that at the black hole horizon timelike particles (that are pulled by gradient of potential produced by matter (non-gravitational) energy) tend to photons that can feel no gravitational pull but only follow curvature of space produced by gravitational field energy. The horizon should therefore be defined when their respective sources are equal in magnitude. That is equipartition between gravitational and non-gravitational (matter) energy. By using \BY energy, gravitational field energy was computed and the equipartition was shown to exist for static black hole horizons. This is the conjecture for location of black hole horizon which we would like to examine for stationary black holes in \gr\ as well as with the appropriate generalization of 
\BY energy for static black holes in \LL gravity. It has  been previously tested for static black holes in \gr \cite{Dadhich:1997ze}. In this work our main motivation, besides obtaining the quasilocal energy for \LL gravity, is to test the veracity of this equipartition conjecture for stationary black holes in \gr\ and static black holes in \LL gravity. 

The paper is organized as follows, in \ref{BY_GR} we discuss the \BY energy in the context of \gr\ and then we have applied it to \RN and Kerr-Newmann black holes to obtain the location of the horizon through equipartition. Then in the subsequent section, i.e., in \ref{BY_LL} we have discussed extensively the quasilocal energy for \LL gravity and equipartition conjecture in it. Finally we conclude with a short discussion on our results. 

In this paper we will work with the $(-,+,+,\ldots)$ signature for the spacetime metric and shall set the fundamental constants $G$ and $c$ to unity. 
\section{Brown-York quasilocal energy and black hole horizons in general relativity}\label{BY_GR}

We will start with a spacetime region $\mathcal{M}$, which is topologically $\Sigma$ times a real line interval, where $\Sigma$ is the three-space. The boundary of the three-space $\Sigma$, is denoted by $\mathcal{B}$ (not necessarily simply connected), which is two dimensional. The product of $\mathcal{B}$ with timelike world lines normal to $\Sigma$ is denoted by $B$ and the end points of the timelike world lines are denoted by $\Sigma _{1}$ and $\Sigma _{2}$. Hence $B$, $\Sigma _{1}$ and $\Sigma _{2}$ form the three-boundary of the full spacetime region $\mathcal{M}$. 

The full spacetime metric is $g_{ab}$, $u_{a}$ is the future pointing timelike normal to the hypersurface $\Sigma$, and $n_{a}$ is the spacelike normal to the three-boundary $B$. The metric and extrinsic curvature on $\Sigma$ are denoted by $h_{ab}$ and $K_{ab}$ respectively, with $h^{a}_{b}$ acting as the projection tensor on $\Sigma$. From hypersurface orthogonality between $\Sigma$ and $B$ it immediately follows that $u_{a}n^{a}=0$. Similar projection tensors and extrinsic curvatures can be defined on $B$ as well. This can be extended to finally introduce the extrinsic curvature $k_{AB}$ on the two-boundary $\mathcal{B}$ of $\Sigma$ with the induced metric, $q^{a}_{b}=\delta ^{a}_{b}+u^{a}u_{b}-n^{a}n_{b}$. Given this spacetime foliation we can introduce the ADM decomposition \cite{Arnowitt:1962hi,gravitation,Poisson} and thus the action can be written as a bulk term which includes intrinsic quantities defined on $\Sigma$ and a surface term on the three-boundary. On the two end points $\Sigma _{1}$ and $\
Sigma _{2}$ the surface term equals $2K$, where $K=K_{ab}h^{ab}$ is the trace of the extrinsic curvature on $\Sigma$. On the other surface, namely $B$, the surface term equals the trace of the extrinsic curvature defined on $B$. Then under variation we will obtain the gravitational momentum conjugate to both $h_{ab}$ on $\Sigma$ and the respective one on $B$. Treating the gravitational action analogously to a matter action and from the Hamilton-Jacobi method the quasilocal gravitational energy contained within the two-surface $\mathcal{B}$ turns out to be \cite{Brown:1992br}
\begin{align}\label{BY_GR_Eq01}
E_{\rm{BY}}=\frac{1}{8\pi}\int _{\mathcal{B}}d^{2}x\sqrt{q}\left(k-k_{0}\right)
\end{align}
where $q$ is the two-metric defined on the two-surface $\mathcal{B}$ and $k_{0}$ stands for the trace of extrinsic curvature for some reference spacetime. As the two-surface $\mathcal{B}$ tends to infinity, the \BY energy would approach the ADM mass (the positivity of \BY energy and its relation to hoop conjecture have been explored in \cite{Shi:2003ab,Fan:2007ab,O'Murchadha:2009kc,Malec:2015oza}). We will be interested in asymptotically flat solutions in which $k_{0}$ is the trace of extrinsic curvature of $\mathcal{B}$ as embedded in a flat spacetime. In this section we will evaluate the quasilocal energy for two cases: (a) static spherically symmetric  and (b) stationary axially symmetric spacetimes, in particular for \RN and  Kerr-Newmann black holes.  In both these cases we will explicitly see that in the asymptotic limit the \BY energy goes to the ADM energy \cite{Arnowitt:1962hi}, which is a crucial check for the validity of any definition of energy in \gr. 
\subsection{\BY quasilocal energy in an arbitrary static spherically symmetric spacetime} 

We will first derive the expression for \BY energy for the most general static spherically symmetric spacetime \cite{Chakraborty:2011uj,chakraborty:2015vla,Chakraborty:2012sd} with the following line element
\begin{align}\label{BY_GR_Eq02}
ds^{2}=-f(r)dt^{2}+\frac{dr^{2}}{g(r)}+r^{2}d\Omega ^{2}
\end{align}
The metric is assumed to be asymptotically flat, i.e., in the limit $r\rightarrow \infty$ both $f(r)$ and $g(r)$ go to unity giving the flat Minkowskian metric.
The hypersurface $\Sigma$ is taken to be a $t=\textrm{constant}$ hypersurface and $B$ as a $r=\textrm{constant}$ surface. The two-surface $\mathcal{B}$ is a $r=\textrm{constant}$ hypersurface within $\Sigma$. Then the unit timelike normal $u_{a}$ to $\Sigma$ and unit spacelike normal $n_{a}$ to $B$ turn out to be 
\begin{align}
u_{a}&=-\sqrt{f}\left(1,0,0,0\right);\qquad u^{a}=\frac{1}{\sqrt{f}}\left(1,0,0,0\right)
\label{BY_GR_Eq03}
\\
n_{a}&=\frac{1}{\sqrt{g}}\left(0,1,0,0\right);\qquad n^{a}=\sqrt{g}\left(0,1,0,0\right)
\label{BY_GR_Eq04}
\end{align}
Having obtained the normal to $\Sigma$ and $B$ we can now compute the corresponding extrinsic curvatures, and in particular for the latter we have,  
\begin{align}\label{BY_GR_Eq05}
k&=-\frac{1}{\sqrt{h}}\partial _{\mu}\left(\sqrt{h}n^{\mu}\right)=-\partial _{r}n^{r}-n^{r}\partial _{r}\ln \sqrt{h}
\nonumber
\\
&=-\partial _{r}\sqrt{g}-\sqrt{g}\partial _{r}\ln \left(\frac{r^{2}\sin \theta}{\sqrt{g}}\right)
=-\frac{2}{r}\sqrt{g(r)}
\end{align}
The embedding of $\mathcal{B}$ in a flat spacetime is trivial and the trace of extrinsic curvature $k_{0}$ is simply $-2/r$. Then using the expression for $k$ from \ref{BY_GR_Eq05} the \BY energy as defined in \ref{BY_GR_Eq01} yields,
\begin{align}\label{BY_GR_Eq06}
E_{\rm{BY}}=\frac{1}{4\pi}\int d\theta d\phi ~r^{2}\sin \theta \frac{1}{r}\left(1-\sqrt{g(r)}\right)
=r\left(1-\sqrt{g(r)}\right)
\end{align}
This expression for \BY energy is completely general. Given any static, spherically symmetric spacetime in \gr\ with line element given by \ref{BY_GR_Eq02} the above result holds. This also shows that the \BY energy singles out the coefficient of $g_{rr}$, rather than that of $g_{tt}$. The implication of which is straightforward --- the gravitational energy really resides in  the spatial curvature, i.e. it curves the three space. This is in complete accord with some of the earlier studies by one of us \cite{Dadhich:1997ku}.

Having derived the general expression, let us now consider an application of this result. For that we pick up the \RN black hole, whose metric elements are given by 
\begin{align}\label{Eq06}
f(r)=g(r)=1-\frac{2M}{r}+\frac{Q^2}{r^2}
\end{align}
The \BY quasilocal energy within a sphere of radius $r_{0}$ turns out to be,
\begin{align}\label{BY_GR_Eq07}
E(r\leq r_{0})=r_{0}\left[1-\sqrt{1-\frac{2M}{r_{0}}+\frac{Q^{2}}{r_{0}^{2}}}\right]
\end{align}
where $M$ and $Q$ respectively stand for mass and charge of the black hole. Let us expand it for large $r$ to write 
\begin{align}\label{BY_Asymp}
E(r\leq r_{0})\approx M-\frac{Q^{2}}{2r_{0}}+\frac{M^{2}}{2r_{0}}
\end{align}
Clearly it goes to the desired limit, the ADM mass M, asymptotically. It is also evident from the expression of quasi-local energy that, E is the sum of matter energy and potential energy associated with building a charged fluid ball by bringing together individual particles from some initial radius. It can also have the following understanding: $M$ being the total energy including rest mass and all kinds of interaction energies. The energy lying exterior to the radius $r_{0}$ will be $Q^{2}/2r_{0}$, arising from the energy momentum tensor component $T^{0}_{0}=Q^{2}/2r_{0}^{4}$ due to electric field. The contribution of gravitational potential energy corresponds to the second term in the approximation, i.e., $-M^{2}/2r_{0}$. Thus the energy within radius R will correspond to $M-(Q^{2}/2r_{0}-M^{2}/2r_{0})$, exactly coinciding with \ref{BY_Asymp}.

For extremal black hole $M=Q$, the two energies cancel out each other exactly, not only in the asymptotic expansion but everywhere. Thus alike the Komar mass for the Schwarzschild black hole, the \BY energy is conserved, and is equal to ADM mass, for the extremal \RN black hole. The energy within radius $r_{0}$ is given by the \BY expression as presented in \ref{BY_GR_Eq07}, while the energy outside $r_{0}$ has two parts --- (a) the energy contained within the gravitational field $E_{\rm grav}$ and (b) the energy contained within the electric field $Q^{2}/2r_{0}$ \cite{Dadhich:1997ze}. This separation of energy into two parts can be written as,
\begin{align}\label{BY_New_01}
E(r\geq r_{0})=E_{\rm{grav}}+\frac{Q^{2}}{2r_{0}}
\end{align}
The total energy in the full spacetime manifold corresponds to the ADM mass. This implies, $E(r\leq r_{0})+E(r\geq r_{0})=M$, the ADM mass. Using which the gravitational energy turns out to be
\begin{align}
E_{\rm{grav}}=M-\frac{Q^{2}}{2r_{0}}-r_{0}\left[1-\sqrt{1-\frac{2M}{r_{0}}+\frac{Q^{2}}{r_{0}^{2}}}\right]
\end{align}
On the other hand the non-gravitational energy outside radius $r_{0}$ arises from the energy density $Q^{2}/2r_{0}^{4}$ due to electric field. Integrating the electric field energy over the range of radial distance from $\infty$ to $r_{0}$, we get
$-Q^{2}/2r_{0}$. Adding M to it we obtain the non-gravitational energy to be, $E_{\rm non-grav}=M-Q^{2}/2r_{0}$. Now according to the  equipartition conjecture, the horizon is defined when energy is equally divided between the matter fields and the gravitational field, leading to, $E_{\rm{grav}}+E_{\rm{non-grav}}=0$, which implies
\begin{align}
\sqrt{1-\frac{2M}{r_{0}}+\frac{Q^{2}}{r_{0}^{2}}}\left[1-\sqrt{1-\frac{2M}{r_{0}}+\frac{Q^{2}}{r_{0}^{2}}}\right]=0.
\end{align}
This clearly has two solutions, namely, (a) $r=r_{+} = M+\sqrt{M^2-Q^2}$, the larger root of  $r^{2}-2Mr+Q^{2}=0$, yielding the location of the event horizon and (b) $r=Q^{2}/2M$, the hard core radius for naked singularity for the parameter space $Q^{2}>M^{2}$. In the latter case, note that the hard core radius marks vanishing of non-gravitational (matter) energy which consequently implies vanishing of gravitational energy as well because the latter is created by the former. Then energy contained inside the hard core radius also vanishes and the entire mass $M$ lies outside. The hard core radius should rather be looked upon as the radius where non-gravitational energy goes to zero rather than equipartition (because it implies zero equal to zero). Thus  \emph{equipartition of energy between gravitational and non-gravitational energy characterizes horizon}.
\subsection{Kerr-Newmann Black Hole}

Following on the \RN black hole, we will now compute the \BY energy for the Kerr-Newman black hole and verify the veracity of the equipartition conjecture for the location of its horizon. It is described by the metric
\begin{align}\label{BY_GR_Eq12}
ds^{2}=-\left(\frac{\Delta -a^{2}\sin ^{2}\theta}{\rho ^{2}}\right)dt^{2}-\frac{2a\sin ^{2}\theta\left(r^{2}+a^{2}-\Delta\right)}{\rho ^{2}}dtd\phi +\frac{\rho ^{2}}{\Delta}dr^{2}+\rho ^{2}d\theta ^{2}+\frac{\Sigma}{\rho ^{2}}\sin ^{2}\theta d\phi ^{2}
\end{align}
where we have defined the following quantities,
\begin{align}\label{BY_GR_Eq13}
\Delta =r^{2}+a^{2}+Q^{2}-2Mr;\qquad \rho ^{2}=r^{2}+a^{2}\cos ^{2}\theta ;
\qquad \Sigma =\left(r^{2}+a^{2}\right)^{2}-\Delta a^{2}\sin ^{2}\theta
\end{align}
where $M$, $Q$ and $a$ have the usual meaning as mass, charge and specific angular momentum of the black hole.  The normalized normal to $r=\textrm{constant}$ surface within the $t=\textrm{constant}$ surface is spacelike, i.e., $n_{a}n^{a}=1$, which for the spacetime described by the above metric ansatz turns out to be,
\begin{align}\label{BY_GR_Eq14}
n_{\mu}=\left(\sqrt{\frac{\rho ^{2}}{\Delta}},0,0\right);\qquad n^{\mu}=\left(\sqrt{\frac{\Delta}{\rho ^{2}}},0,0\right)
\end{align}
Thus the extrinsic curvature of any $r=\textrm{constant}$ surface within the $t=\textrm{constant}$ hypersurface can be obtained as
\begin{align}\label{BY_GR_Eq15}
k=-\frac{1}{\sqrt{h}}\partial _{\mu}\left(\sqrt{h}n^{\mu}\right)=-\partial _{r}n^{r}-n^{r}\partial _{r}\ln \sqrt{h}
\end{align}
The above expression nicely breaks into two parts, $\partial _{r}n^{r}$ and $\partial _{r}\ln \sqrt{h}$. Each of them can be evaluated individually leading to
\begin{align}\label{BY_GR_Eq16}
\partial _{r}n^{r}=\frac{1}{2}\sqrt{\frac{\Delta}{\rho ^{2}}}\left(\frac{\partial _{r}\Delta}{\Delta}-\frac{\partial _{r}\rho ^{2}}{\rho ^{2}}\right);\qquad
\partial _{r}\ln \sqrt{h}=\frac{\partial _{r}\rho ^{2}}{2\rho ^{2}}+\frac{\partial _{r}\Sigma}{2\Sigma}-\frac{\partial _{r}\Delta}{2\Delta}
\end{align}
This immediately leads to the following expression for extrinsic curvature
\begin{align}\label{BY_GR_Eq17}
k&=-\sqrt{\frac{\Delta}{\rho ^{2}}}\frac{\partial _{r}\Sigma}{2\Sigma}
\nonumber
\\
&=-r\sqrt{1-\frac{2M}{r}+\frac{a^{2}+Q^{2}}{r^{2}}}\frac{2r\left(r^{2}+a^{2}\right)-\left(r-M\right)a^{2}\sin ^{2}\theta}{\sqrt{r^{2}+a^{2}\cos ^{2}\theta}\left[\left(r^{2}+a^{2}\right)^{2}-a^{2}\Delta \sin ^{2}\theta \right]}
\end{align}
Now the metric on the two-surface is $\sqrt{q}=\sqrt{\Sigma}\sin \theta$. Hence the unreferenced \BY energy for the Kerr-Newmann black hole within a sphere of radius $r$ turns out to be
\begin{align}\label{BY_GR_Eq18}
\mathcal{E}&=\frac{1}{8\pi}\int d^{2}x \sqrt{q}~k=\frac{1}{8\pi}\int d\phi d\theta \sin \theta \sqrt{\Sigma}k
\nonumber
\\
&=-\frac{1}{4}r\sqrt{1-\frac{2M}{r}+\frac{a^{2}+Q^{2}}{r^{2}}}\int _{0}^{\pi}d\theta \sin \theta \frac{2r\left(r^{2}+a^{2}\right)-\left(r-M\right)a^{2}\sin ^{2}\theta}{\sqrt{\left(r^{2}+a^{2}\cos ^{2}\theta\right)\left[\left(r^{2}+a^{2}\right)^{2}-a^{2}\Delta \sin ^{2}\theta \right]}}
\end{align}
which in general cannot be integrated to obtain a closed form expression. However a closed form expression is indeed possible to obtain in the slow rotation limit i.e., $a/r\ll 1$. Then the above expression for unreferenced \BY energy reduces to \cite{Bose:1999er}
\begin{align}\label{BY_GR_Eq19}
\mathcal{E}&=-\frac{1}{2}r\sqrt{1-\frac{2M}{r}+\frac{a^{2}+Q^{2}}{r^{2}}}\int _{0}^{\pi}d\theta \sin \theta \left[1-\frac{a^{2}}{2r^{2}}\left\lbrace \cos ^{2}\theta +\left(\frac{M}{r}-\frac{Q^{2}}{r^{2}}\right)\sin ^{2}\theta \right\rbrace\right]
\nonumber
\\
&=-r\sqrt{1-\frac{2M}{r}+\frac{a^{2}+Q^{2}}{r^{2}}}\left[1-\frac{a^{2}}{6r^{2}}\left(1+\frac{2M}{r}-\frac{Q^{2}}{r^{2}}\right)\right]
\end{align}
Getting the reference term is more difficult, for that we have to map the two dimensional surface to a flat three dimensional spacetime. If the flat three dimensional surface is described by the three coordinates $R,\vartheta ,\Phi$, then the matching with two-surface would provide the following relations, $R=R(\theta), \vartheta =\vartheta (\theta)$ and $\Phi =\phi$. These relations have to be obtained through their substitution in the flat space line element and comparing with two-surface and Kerr-Newmann metric. This leads to the following differential equations for $R$ and $\vartheta$ 
\begin{align}
\left(\dfrac{dR}{d\theta}\right)^{2}&+R^{2}\left(\frac{d\vartheta}{d\theta}\right)^{2}=\rho ^{2}
\label{BY_GR_Eq20a}
\\
R^{2}\sin ^{2}\vartheta&=\frac{\Sigma}{\rho ^{2}}\sin ^{2}\theta
\label{BY_GR_Eq20b}
\end{align}
We need to solve these two coupled differential equations to get both $R$ and $\vartheta$ in terms of $\theta$. \ref{BY_GR_Eq20b} can be solved to get $R$ in terms of $\vartheta$ and $\theta$. This when substituted in \ref{BY_GR_Eq20a} would yield a differential equation of $\vartheta$. However the differential equation being complicated, in general (i.e., for arbitrary choices of the rotation parameter $a$) would not posses any analytic closed form solution. Thus to get analytic expression we need to use slow rotation limit, in which the solutions for $R(\theta)$ and $\vartheta (\theta)$ can be obtained as
\begin{align}
\sin \vartheta &=\sin \theta \left[1+\frac{a^{2}}{2r^{2}}\left(1+\frac{2M}{r}-\frac{Q^{2}}{r^{2}}\right)\cos ^{2}\theta \right]
\label{BY_GR_Eq21a}
\\
R(\theta)&=r\left[1+\frac{a^{2}}{2r^{2}}\sin ^{2}\theta -\frac{a^{2}}{2r^{2}}\left(\frac{2M}{r}-\frac{Q^{2}}{r^{2}} \right)\cos ^{2}\theta \right]
\label{BY_GR_Eq21b}
\end{align}
From these two relations we could obtain an expression for the extrinsic curvature $k_{0}$ of a two surface as embedded in flat space. Then this extrinsic curvature can be used to compute the reference term, which yields \cite{Bose:1999er}
\begin{align}\label{BY_New_Ref}
\mathcal{E}_{0}=-r_{0}\left[1+\frac{a^{2}}{3r_{0}^{2}}\left(1+\frac{M}{r_{0}}-\frac{Q^{2}}{2r_{0}^{2}}\right)\right]
\end{align}
It is interesting to note that there also appears the contribution of rotational energy in the flat space referenced energy. On  expansion of $\mathcal{E}_{0}$ in powers of $1/r$,  the first order correction due to rotational energy is  $-a^{2}/3r_{0}$, we will  comment on it  later on.

Using the expression for $\mathcal{E}$ from \ref{BY_GR_Eq19} and the reference term as in \ref{BY_New_Ref}, the referenced \BY energy turns out to be \cite{Bose:1999er}
\begin{align}\label{BY_GR_Eq22}
E(r\leq r_{0})\equiv \mathcal{E}-\mathcal{E}_{0}=r_{0}\left[1-\sqrt{1-\frac{2M}{r_{0}}+\frac{a^{2}+Q^{2}}{r_{0}^{2}}}\right] &+\frac{a^{2}}{6r_{0}}\Big[2\left(1+\frac{M}{r_{0}}-\frac{Q^{2}}{2r_{0}^{2}}\right)
\nonumber
\\
&+\left(1+\frac{2M}{r_{0}}-\frac{Q^{2}}{r_{0}^{2}}\right)\sqrt{1-\frac{2M}{r_{0}}+\frac{Q^{2}}{r_{0}^{2}}}\Big]
\end{align}
which gives the  quasilocal energy within a sphere of radius $r_{0}$. If we had taken the sphere of radius $r_{0}$ to infinity, it would as expected go to  the ADM mass $M$. Expanding terms within square root in powers of $1/r$ we arrive at 
\begin{align}
E(r\leq r_{0})\approx M-\frac{a^{2}+Q^{2}-M^{2}}{2r_{0}}-\frac{M(a^{2}+Q^{2})}{2r_{0}^{2}}+\frac{a^{2}}{6r_{0}}\Big[3+\frac{3M}{r_{0}}-\frac{2M^{2}}{r_{0}^{2}}-2\frac{Q^{2}}{r_{0}^{2}}+\frac{Q^{2}-M^{2}}{2r_{0}^{2}}\Big]
\end{align}
Note that there is no contribution of rotational energy at order $1/r_{0}$. This is precisely because rotational energy along with Kerr-Newmann spacetime, is also shared by the referenced spacetime. Thus as the reference term $\mathcal{E}_{0}$ is subtracted in order to get the \BY energy, the contribution from rotational part is exactly canceled. That is why the \BY energy for Kerr-Newmann spacetime is free of pure rotational terms and rotation only contributes through coupling with mass and charge. This is also the reason behind the fact that for the case of extremal black hole (i.e., $M^{2}=a^{2}+Q^{2}$) the \BY energy reduces to $M+(a^{2}/2r_{0})$. Thus in addition to the ADM mass the rotational energy $a^{2}/2r_{0}$ also contributes to the \BY energy at large distance for extremal Kerr-Newmann black hole. Note that for $a=0$, we get back the result for extremal \RN black hole. 

Now let's turn to our main aim of obtaining the location of horizon at the equipartition of gravitational and non-gravitational energy. The  energy outside $r_{0}$ has the expression, $E(r\geq r_{0})=E_{\rm{grav}}+(Q^{2}/2r_{0})$. Since total energy in the spacetime is the ADM mass $M$, we obtain the gravitational energy to be
\begin{align}
E_{\rm{grav}}=M-\frac{Q^{2}}{2r_{0}}-r_{0}\left[1-\sqrt{1-\frac{2M}{r_{0}}+\frac{a^{2}+Q^{2}}{r_{0}^{2}}}\right] &-\frac{a^{2}}{6r_{0}}\Bigg[2\left(1+\frac{M}{r_{0}}-\frac{Q^{2}}{2r_{0}^{2}}\right)
\nonumber
\\
&+\left(1+\frac{2M}{r_{0}}-\frac{Q^{2}}{r_{0}^{2}}\right)\sqrt{1-\frac{2M}{r_{0}}+\frac{Q^{2}}{r_{0}^{2}}}\Bigg]
\end{align}
On the other hand the non-gravitational energy is $E_{\rm{non-grav}}=M-(a^{2}+Q^{2}/2r_{0})$. Requiring  equipartition of the two,  i.e., gravitational and non-gravitational to be equal (with proper sign to ensure attractive nature of gravity), we finally obtain
\begin{align}
\sqrt{1-\frac{2M}{r_{0}}+\frac{a^{2}+Q^{2}}{r_{0}^{2}}}\Bigg[\sqrt{1-\frac{2M}{r_{0}}+\frac{a^{2}+Q^{2}}{r_{0}^{2}}}-1&-\frac{a^{2}}{6r_{0}^{2}}\sqrt{1-\frac{2M}{r_{0}}+\frac{a^{2}+Q^{2}}{r_{0}^{2}}}
\nonumber
\\
&+\frac{a^{2}}{6r_{0}^{2}}\left(1+\frac{2M}{r_{0}}-\frac{Q^{2}}{r_{0}^{2}}\right)\Bigg]=0
\end{align}
This algebraic equation has two solutions: (a) $r=r_{+}=M+\sqrt{M^2-a^2-Q^2}$, the larger root of $r^{2}-2Mr+(Q^{2}+a^{2})=0$ defining the horizon, and (b) $r=(a^{2}+Q^{2})/2M$, the hard core radius for naked singularity, $a^{2}+Q^{2}>M^2$. Thus the equipartition conjecture is also verified for the Kerr-Newman black hole albeit in the slow rotation limit. 

Having succeeded in deriving the horizon (or hard radius) location, starting from equipartition in the case of both charged and charged axi-symmetric configurations in \gr\ it would be interesting to ask, what happens in higher dimensions, still in the premise of \gr\ or more importantly, what happens to this equipartition conjecture when higher order curvature invariants are present in the action. That is what we take up  in the next section.
\section{\BY quasilocal energy and Black Hole Horizons in \LL Gravity}\label{BY_LL}

Before addressing the \BY energy for \LL gravity, as a warm up as well as to show the complexities involved, we will first discuss the case of \BY energy in higher dimension but within the context of \gr. The \BY energy for a D-dimensional spacetime in \gr\ is still given by the difference $k-k_{0}$ but this time evaluated for a $(D-2)$-dimensional surface. The geometry remains similar, the full manifold $\mathcal{M}$ is $D$-dimensional, in which we have constant time hypersurfaces, namely $\Sigma$, which are now $(D-1)$-dimensional and so is the surface $B$. Hence the boundary of $\Sigma$ as intersected by $B$ forms the desired co-dimension two-surface $\mathcal{B}$, which is now $(D-2)$-dimensional. However we need to account for proper volume factors. For this reason the \BY energy is written as
\begin{align}\label{BY_LL_Eq01}
E_{BY}=\frac{1}{8\pi \gamma}\int dA_{D-2}\left(k-k_{0}\right)
\end{align}
where $\gamma$ is a numerical factor to be fixed later. Any static spherically symmetric spacetime can be presented by the metric   \ref{BY_GR_Eq02}. The extrinsic curvature of the $(D-2)$-dimensional surface $\mathcal{B}$ is $k=-[(D-2)/r]\sqrt{g(r)}$ and $k_{0}=-(D-2)/r$. On substitution in \ref{BY_LL_Eq01} and integration over the $(D-2)$-dimensional area immediately yields
\begin{align}\label{BY_LL_Eq02}
E_{BY}=\frac{D-2}{8\pi \gamma}\frac{2\pi ^{(D-1)/2}}{\Gamma (\frac{D-1}{2})}r^{D-3}\left(1-\sqrt{g(r)}\right)
\end{align}
By setting the constant $\gamma$ to the value
\begin{align}\label{BY_LL_Eq03}
\gamma=\frac{D-2}{\Gamma (\frac{D-1}{2})}\frac{\pi ^{(D-3)/2}}{4}
\end{align}
we obtain the following expression for \BY energy in a static spherically symmetric spacetime in \gr\ in higher dimension
\begin{align}\label{BY_LL_Eq04}
E_{BY}=r^{D-3}\left(1-\sqrt{g(r)}\right)
\end{align}
In $D$-dimensions static spherically symmetric solution of \gr\ corresponds to, $f=g^{-1}=1-(M/r^{D-3})$, which when substituted in the above expression, leads to $\lim _{r\rightarrow \infty}E_{BY}=M$, i.e., the above expression agrees as required  with ADM mass in the asymptotic limit. 

However we cannot use this for \LL gravity. This can be seen directly as in $m$th order \LL gravity the  static  spherically symmetric  black hole solution is given by  $f(r)=g(r)=1-(\gamma /r^{(D-(2m+1))/m})$. In analogy with the Einstein case we write from \ref{BY_LL_Eq04} the \BY energy as 
\begin{align}\label{BY_LL_Eq05}
E_{BY}=r^{{\frac{D-2m-1}{m}}}\left[1-(1-\frac{\gamma}{r^{\frac{D-2m-1}{m}}})^{1/2}\right]\stackrel{r\rightarrow \infty}{=}\frac{\gamma}{2}
\end{align}
which for large $r$ goes as $\gamma/2$. It does not agree with the ADM mass which should really be $\propto \gamma^m$ \cite{Kastor:2008xb}. However it is interesting that we get almost the right result but for mismatch in dimension of mass. This is of course not the right expression for quasi-local \BY energy, and we need to re-derive the \BY energy expression which would be appropriate for the \LL theory. That is what we will do next.   
\subsection{\BY quasilocal energy in \LL Gravity: Formalism}

The \BY energy for \LL gravity can be obtained in a straightforward manner following the \gr\ prescription. The derivation essentially amounts to calculate the counter term for gravitational action on a $t=\textrm{constant}$ surface (e.g., $\Sigma$) and the corresponding term for the $(D-2)$-boundary $\mathcal{B}$. Thus we can use the counter term for \LL gravity \cite{Kofinas:2007ns,Miskovic:2007mg,Barrabes:2005ar,Brihaye:2008xu,Brihaye:2013vsa,Brihaye:2010wx,Padmanabhan:2013xyr} on the $t=\textrm{constant}$ hypersurface $\Sigma$ and then take it to the $(D-2)$-boundary $\mathcal{B}$. This leads to the following generalization of \BY energy to $m$th order pure \LL gravity as,
\begin{align}\label{BY_LL_Eq06}
E_{BY}^{(m)}=c^{(m)}\int dA_{D-2}\mathcal{E}
\end{align}
where $c^{(m)}$ is a numerical factor which can be adjusted to get $E_{BY}=M$ at infinity. In \gr\ the quantity $\mathcal{E}$ was just $(k-k_{0})$ and $c^{(1)}=(\Gamma (\frac{D-1}{2}))/(2(D-2)\pi ^{(D-1)/2})$, however for $m$th order \LL gravity $\mathcal{E}$ turns out to be
\begin{align}\label{BY_LL_Eq07}
\mathcal{E}=\frac{m!}{2^{m+1}}\sum _{s=0}^{m-1}c_{s}\Pi ^{(s)}
\end{align}
where,
\begin{align}\label{BY_LL_Eq08}
\Pi ^{(s)}=\delta ^{A_{1}A_{2}\ldots A_{2m-1}}_{B_{1}B_{2}\ldots B_{2m-1}}R^{B_{1}B_{2}}_{A_{1}A_{2}}\cdots R^{B_{2s-1}B_{2s}}_{A_{2s-1}A_{2s}}\left(k^{B_{2s+1}}_{A_{2s+1}}-k^{B_{2s+1}}_{(0)A_{2s+1}} \right)\cdots \left(k^{B_{2m-1}}_{A_{2m-1}}-k_{(0)A_{2m-1}}^{B_{2m-1}}\right)
\end{align}
Here $k_{(0)B}^{A}$ is the extrinsic curvature of the two surface as embedded in flat spacetime. Also the arbitrary constants $c_{s}$ appearing in \ref{BY_LL_Eq07} has the following expression:
\begin{align}\label{BY_LL_New}
c_{s}=\sum _{q=s}^{m-1}\frac{(-2)^{q-s}4^{m-q}~(~^{q}C_{s})}{q!(2m-2q-1)!!}
\end{align}
Here$~^{q}C_{s}$ is the combination symbol and stands for $q!/\{(q-s)!s!\}$. Before proceeding further let us pause for a while, and consider the case  of static vacuum solution for $m$th order pure \LL gravity which is given by  \cite{Crisostomo:2000bb,Deser:2002rt,Baykal:2012rr,Kastor:2008xb} 
\begin{align}\label{BY_LL_Eq09}
ds^{2}=-f(r)dt^{2}+\frac{dr^{2}}{f(r)}+r^{2}d\Omega _{D-2}^{2};\qquad f(r)=1-\left(\beta/r^{\frac{D-(2m+1)}{m}}\right)
\end{align}
It is clear that $\beta$ is a dimension full constant related to the ADM mass and $D$ is the dimension of the spacetime such that $D\geq 2m+2$. Note that $D\geq 2m+1$ ensures that the Lagrangian is not a topological term and $D=2m+1$ is excluded because it represents a solid angle deficit that can only describe a global monopole \cite{Barriola:1989} without black hole. The later computations would involve the curvature tensor components which are 
\begin{align}\label{BY_LL_Eq10}
R^{kl}_{ij}=\frac{1}{r^{2}}(1-f)\delta ^{kl}_{ij}
\end{align}
However the ADM mass defined in the context of $m$th order \LL gravity turns out to be $M=\hat{c}\beta ^{m}$, where $\hat{c}$ is just a numerical constant. This numerical constant depends explicitly on the order of \LL Lagrangian and also on the coefficient of m-th \LL Lagrangian, $c^{(m)}$. The exact value of $\hat{c}$ in this and subsequent examples would not hit us, since we will be concerned primarily with the structure of the results. Thus the acid test for our prescription of \BY energy as presented in \ref{BY_LL_Eq06} would be to see whether it matches with the ADM mass $M$ as defined above.

In order to show the validity of our result, let us first see that it includes the \gr\ case for $m=1$. We evaluate \ref{BY_LL_Eq06} for $m=1$, which immediately leads to $s=0$. Using this in \ref{BY_LL_Eq07} and \ref{BY_LL_Eq08} we readily obtain, \ref{BY_LL_Eq01}, the correct quasilocal energy for \gr. 

Next we take up the Gauss-Bonnet action, which corresponds to $m=2$. Thus we have two possibilities $s=1$ and $s=0$ respectively. For $s=0$, we obtain
\begin{align}
\Pi ^{(0)}=\delta ^{A_{1}A_{2}A_{3}}_{B_{1}B_{2}B_{3}}\left(k^{B_{1}}_{A_{1}}-k_{(0)A_{2}}^{B_{2}}\right)\left(k^{B_{2}}_{A_{2}}-k_{(0)A_{2}}^{B_{2}}\right)\left(k^{B_{3}}_{A_{3}}-k_{(0)A_{3}}^{B_{3}}\right)\sim \frac{\beta ^{3}}{r^{3}r^{3(D-5)/2}}
\end{align}
while for $s=1$, we get
\begin{align}
\Pi ^{(1)}=\delta ^{A_{1}A_{2}A_{3}}_{B_{1}B_{2}B_{3}}R^{B_{1}B_{2}}_{A_{1}A_{2}}\left(k^{B_{3}}_{A_{3}}-k_{(0)A_{3}}^{B_{3}}\right)\sim \frac{1}{r^{3}}\frac{\beta ^{2}}{r^{D-5}}
\end{align}
Then the \BY energy at large $r$ takes the form 
\begin{align}
E_{BY}&=c^{(2)}r^{D-2}\left[c_{0}\frac{\beta ^{3}}{r^{3}r^{3(D-5)/2}}+c_{1}\frac{1}{r^{3}}\frac{\beta ^{2}}{r^{D-5}}\right]
\nonumber
\\
&=c^{(2)}c_{1}\beta ^{2}+c^{(2)}c_{0}\beta ^{3}r^{-\frac{D-5}{2}}\stackrel{r\rightarrow \infty}{=} \hat{c}\beta ^{2}=M
\end{align}
which exactly agrees  with the ADM mass. Hence for pure Gauss-Bonnet gravity, the \BY energy at infinity exactly matches with the ADM mass. 

It is now time to consider $m$th order pure \LL gravity in which $s$ can take values $0,1,\ldots ,(m-1)$ respectively. This immediately leads to the following expression
\begin{align}
\Pi ^{(0)}&=\delta ^{A_{1}A_{2}\ldots A_{2m-1}}_{B_{1}B_{2}\ldots B_{2m-1}}\left(k^{B_{1}}_{A_{1}}-k_{(0)A_{2}}^{B_{2}}\right)\left(k^{B_{2}}_{A_{2}}-k_{(0)A_{2}}^{B_{2}}\right)\cdots \left(k^{B_{2m-1}}_{A_{2m-1}}-k_{(0)A_{2m-1}}^{B_{2m-1}}\right)
\nonumber
\\
& \sim \frac{1}{r^{2m-1}}\left(1-\sqrt{f}\right)^{2m-1} \xrightarrow{r\rightarrow \infty}  \frac{1}{r^{2m-1}}\frac{\beta ^{2m-1}}{r^{(2m-1)(D-2m-1)/m}}
\end{align}
Then,
\begin{align}
\Pi ^{(s)}&=\delta ^{A_{1}A_{2}\ldots A_{2m-1}}_{B_{1}B_{2}\ldots B_{2m-1}}R^{B_{1}B_{2}}_{A_{1}A_{2}}\cdots R^{B_{2s-1}B_{2s}}_{A_{2s-1}A_{2s}}\left(k^{B_{2s+1}}_{A_{2s+1}}-k^{B_{2s+1}}_{(0)A_{2s+1}} \right)\cdots \left(k^{B_{2m-1}}_{A_{2m-1}}-k_{(0)A_{2m-1}}^{B_{2m-1}}\right)
\nonumber
\\
&\sim \frac{1}{r^{2m-1}}\left(1-f\right)^{s}\left(1-\sqrt{f}\right)^{2m-2s-1} \xrightarrow{r\rightarrow \infty} \frac{1}{r^{2m-1}}\frac{\beta ^{2m-s-1}}{r^{(2m-s-1)(D-2m-1)/m}}
\end{align}
and finally,
\begin{align}
\Pi ^{(m-1)}&=\delta ^{A_{1}A_{2}\ldots A_{2m-1}}_{B_{1}B_{2}\ldots B_{2m-1}}R^{B_{1}B_{2}}_{A_{1}A_{2}}\cdots R^{B_{2m-3}B_{2m-2}}_{A_{2m-3}A_{2m-2}}\left(k^{B_{2m-1}}_{A_{2m-1}}-k_{(0)A_{2m-1}}^{B_{2m-1}}\right)
\nonumber
\\
&\sim \frac{1}{r^{2m-1}}\left(1-f\right)^{m-1}\left(1-\sqrt{f}\right)\xrightarrow{r\rightarrow \infty} \frac{1}{r^{2m-1}}\frac{\beta ^{m}}{r^{D-2m-1}}
\end{align}
Hence the  \BY energy at large $r$  turns out to have the following expression
\begin{align}
\lim _{r\rightarrow \infty}E_{BY}&=c^{(m)}r^{D-2}\Big[c_{(0)}\frac{1}{r^{2m-1}}\frac{\beta ^{2m-1}}{r^{(2m-1)(D-2m-1)/m}}+\cdots +c_{(s)}\frac{1}{r^{2m-1}}\frac{\beta ^{2m-s-1}}{r^{(2m-s-1)(D-2m-1)/m}}
\nonumber
\\
&+\cdots +c_{(m-1)}\frac{1}{r^{2m-1}}\frac{\beta ^{m}}{r^{D-2m-1}}\Big]
\nonumber
\\
&=\hat{c}\beta ^{m}+\cdots +\bar{c}_{(s)}\left(\beta ^{2m-s-1}/r^{(D-2m-1)(m-s-1)/m} \right)
\nonumber
\\
&+\cdots +\bar{c}_{(0)}\left(\beta ^{2m-1}/r^{(D-2m-1)(m-1)/m}\right)\stackrel{r\rightarrow \infty}{=} M 
\end{align}
Thus  we have proved that the \BY energy as defined by \ref{BY_LL_Eq06} in the asymptotic limit leads to the ADM mass. Hence the definition of \BY energy works perfectly well and it passes the acid test of matching with ADM mass asymptotically for static black hole  in \LL gravity. We have thus generalized the \BY energy for \LL gravity.
\subsection{\BY quasilocal energy in \LL gravity and equipartition}

Having derived the quasilocal energy for \LL theories of gravity, we would now like to apply the equipartition conjecture to \LL gravity and verify its veracity. For any static spherically symmetric spacetime given by \ref{BY_LL_Eq09} describing an $m$th order Lovelock static black hole, the \BY energy reads as   
\begin{align}\label{BY_New_02}
E_{\rm{BY}}&=r^{D-2m-1}\Big[\bar{c}_{(0)}\left(1-\sqrt{f}\right)^{2m-1}+\cdots +\bar{c}_{(s)}\left(1-\sqrt{f}\right)^{2m-2s-1}\left(1-f\right)^{s}
\nonumber
\\
&+\cdots +\bar{c}_{m-1}\left(1-f\right)^{m-1}\left(1-\sqrt{f}\right)\Big]
\end{align}
This is the energy, $E_{\rm{BY}}(r\leq r_{0})$, lying inside radius $r=r_{0}$. Let us consider a charged black hole with   
\begin{align}
f(r)=1-\left(\frac{\beta ^{m}}{r^{D-2m-1}}-\frac{\bar{Q}^{2}}{r^{2D-2m-4}}\right)^{1/m}
\end{align}
where $\beta ^{m} =2M/\bar{c}$ and $\bar{c}=\bar{c}_{(0)}+\cdots +\bar{c}_{(s)}+\cdots +\bar{c}_{(m)}$. For Maxwell field the energy outside $r=r_{0}$ is as before $E(r\geq r_{0})=E_{\rm{grav}}+(Q^{2}/2r_{0}^{D-3})$ while non-gravitational component is $E_{\rm{non-grav}}=M-(Q^{2}/2r_{0}^{D-3})$. Requiring the total energy in spacetime to be equal to the ADM mass, we readily obtain the gravitational contribution to be,
\begin{align}
E_{\rm{grav}}=M-\frac{Q^{2}}{2r_{0}^{D-3}}-r_{0}^{D-2m-1}\Bigg[\bar{c}_{(0)}\left(1-\sqrt{f_{0}}\right)^{2m-1}&+\cdots +\bar{c}_{(s)}\left(1-\sqrt{f_{0}}\right)^{2m-2s-1}\left(1-f_{0}\right)^{s}
\nonumber
\\
&+\cdots +\bar{c}_{m-1}\left(1-f_{0}\right)^{m-1}\left(1-\sqrt{f_{0}}\right)\Bigg]
\end{align}
where $f_{0}=f(r=r_{0})$. Now the equipartition of gravitational and non-gravitational energy leads to the following algebraic equation
\begin{align}
1&-f_{0}
\nonumber
\\
&=\Bigg[\frac{\bar{c}_{(0)}\left(1-\sqrt{f_{0}}\right)^{2m-1}+\cdots +\bar{c}_{(s)}\left(1-\sqrt{f_{0}}\right)^{2m-2s-1}\left(1-f_{0}\right)^{s}+\cdots +\bar{c}_{m-1}\left(1-f_{0}\right)^{m-1}\left(1-\sqrt{f_{0}}\right)}{\bar{c}_{(0)}+\cdots +\bar{c}_{(s)}+\cdots +\bar{c}_{(m)}}\Bigg]^{1/m}
\label{BY_Final}
\end{align}
Note that the above equation has various powers of $f_{0}$ upto order m, which is also the order of pure \LL Lagrangian. In general this is a complicated algebraic equation to solve for $f_{0}$. However we will avoid this difficulty by performing the following trick: we will substitute $f_{0}=0$ and $f_{0}=1$ in the above equation and see whether it is satisfied. As before it turns out that the above two indeed satisfy \ref{BY_Final}. These two conditions have the two familiar solutions, one defining the horizon $r_{+}$, the larger root of  $r^{2D-2m-4}-\beta ^{m}r^{D-3}+\bar{Q}^{2}=0$ and the other, the hard core radius $r=(\bar{Q}^{2}/\beta ^{m})^{1/D-3}$ for naked singularity, with $M^2<a^2+Q^2$. This is for the pure Lovelock analogue of \RN black hole which includes for $Q=0$ the pure Lovelock analogue of the Schwarzschild black hole.  Thus we verify that the equipartition conjecture continues to hold good for pure \LL static black holes. 

The important point to note is that the conjecture turns out to hold good only for pure Lovelock (for a fixed $m$) black holes but not for Einstein-Lovelock (with sum over $m$) black holes. This is what we show next for the case of Einstein-Gauss-Bonnet black hole. 
\subsection{Equipartition conjecture in Einstein-Gauss-Bonnet gravity}

In the previous section we have concentrated on the pure Lovelock theories and the validity of equipartition conjecture. In this section we will illustrate that Equipartition conjecture does not hold for the general Lovelock theories. For that we will use an action, which is sum of the Einstein-Hilbert and Gauss-Bonnet terms. The static black hole solution corresponding to the Einstein-Gauss-Bonnet action, is known as the Boulware-Deser solution \cite{Boulware:1985wk,Wheeler:1985nh}, it reads as follows:  
\begin{align}
ds^{2}=-f(r)dt^{2}+\frac{dr^{2}}{f(r)}+r^{2}d\Omega _{3}^{2};\qquad f(r)=1-\frac{r^{2}}{2\alpha}\left[-1\pm \sqrt{1+\frac{4\alpha M}{r^{4}}}\right]
\end{align}
Here $\alpha$ is the Gauss-Bonnet coupling constant and $M$ is the mass term, and both of them are of dimension $L^{2}$. There are two branches of the solution, the one with $+$ve sign has the right Einstein limit with attractive gravity while the other (i.e., the one with $-$ve sign) is repulsive. We would therefore choose the former. At $r=0$ we have a curvature singularity, which is cloaked by an event horizon for the $+$ve branch for $M\geq\alpha$ and is located at, $r^{2}=r_{h}^{2}=M-\alpha$. Otherwise the above solution would represent a naked singularity. Let us now compute the \BY energy for the Einstein-Gauss-Bonnet solution. For which the quasi-local energy for $D=5$ involves both $m=1$, the \EH term and the $m=2$, the Gauss-Bonnet term. For this, from \ref{BY_LL_Eq04} and \ref{BY_New_02} we write  
\begin{align}\label{BY_EGB_Ener}
E(r\leq r_{0})=r_{0}^{2}\left(1-\sqrt{f_{0}}\right)+\alpha\left[c_{1}\left(1-\sqrt{f_{0}}\right)^{3}+c_{2}\left(1-f_{0}\right)\left(1-\sqrt{f_{0}}\right)\right]
\end{align}
where $c_{1}$ and $c_{2}$ are two constants whose values can be obtained from \ref{BY_LL_New} and $f_{0}=f(r=r_0)$. As an illustration we can consider the asymptotic limit of the \BY energy defined in \ref{BY_EGB_Ener}. For which the first term yields $r_{0}^{2}\times (M/r_{0}^{2})$, while the second term leads to, $M^{3}/r_{0}^{6}+M^{2}/r_{0}^{4}$. Thus in the asymptotic limit, i.e., $r_{0}\rightarrow \infty$ \BY energy exactly equals the ADM mass $M$.  

Given the \BY energy, the gravitational energy is just the difference, $M-E(r\leq r_{0})$, and non-gravitational energy is anyway the ADM mass $M$. Then the  equipartition demands 
\begin{align}\label{BY_NoWork}
2M=r_{0}^{2}\left(1-\sqrt{f_{0}}\right)+\alpha\left[c_{1}\left(1-\sqrt{f_{0}}\right)^{3}+c_{2}\left(1-f_{0}\right)\left(1-\sqrt{f_{0}}\right)\right]
\end{align}
Clearly this does not define horizon for the Einstein-Gauss-Bonnet black hole. For horizon $f(r_{0})=0$ which gives $2M=r_{h}^{2}+\alpha (c_{1}+c_{2})$. This cannot be satisfied because $c_{1}, c_{2}$ are numerical factors. The equipartition conjecture therefore does not work for Einstein-Gauss-Bonnet and in general for Einstein-Lovelock black holes. It works only for pure Lovelock black holes.  

This explicitly shows the special status of pure \LL gravity as also exposed in \cite{Dadhich:2015lra,Dadhich:2012cv,Dadhich:2012pd}. In this case as well \emph{only} for pure Lovelock black holes, the equipartition conjecture holds good and defines the horizon. Thus equipartition conjecture discerns  the pure \LL gravity from the general Einstein-\LL gravity. 
\section{Discussion}

One of the promising candidates for obtaining quasilocal energy in \gr\ is the \BY energy. This definition is based on the Hamilton-Jacobi treatment of the gravitational action written in terms of ADM variables and then identifying the correct expression for energy. The salient feature of this prescription is that, at asymptotic infinity it correctly reproduces the ADM mass, which acts as an acid test for any definition of energy in \gr. In this work we have generalized the  \BY energy for \LL gravity and have employed it to verify the veracity of the equipartition conjecture for defining  black hole horizon. That is horizon marks equality of gravitational and non-gravitational energy. 

It is envisioned that when a configuration is infinitely dispersed, it has energy equal to ADM mass $M$ as it begins collapsing under its own gravity, it picks up gravitational energy which is negative, and electric field and rotational energy for a charged and rotating black hole. Thus at any finite $r$, there are gravitational and non-gravitational components of energy. The \BY energy gives energy contained inside a given radius, from which if we subtract non-gravitational part, we can compute gravitational energy. This could be done for static black holes and for axially symmetric \KN black hole it can be evaluated in the slow rotation limit. By means of the \BY energy expression, we can tame the notorious gravitational field energy to obtain a quantitative expression. The interesting application of which was made by one of us in proposing the equipartition conjecture \cite{Dadhich:1997ze} for characterization of black hole horizon by equality of gravitational and non-gravitational energy. It is motivated 
by the fact that as horizon is approached timelike particles tend to null particles. Motion of the former is governed by $\nabla\Phi$ produced by non-gravitational energy while that of the latter by spatial curvature caused by gravitational energy \cite{Dadhich:1997ze}. Thus as the former approaches the latter at the horizon so should be their sources. Thus gravitational and non-gravitational energy must be equal at the horizon. 

We have verified the equipartition conjecture for static and axially symmetric (in the slow rotation limit) black holes, in particular \KN black hole that includes static black holes. It is remarkable that for extremal \RN black hole gravitational and electric field energy exactly cancel out each-other everywhere so that the \BY energy is conserved mass $M$. 

It turns out that the definition of  \BY energy cannot straightway be taken over to \LL gravity but it needs to be supplemented with the counter terms. With this modification, we generalize the \BY energy expression for \LL gravity and again separate out gravitational and non-gravitational parts. We use that to establish the equipartition conjecture for static  pure Lovelock black holes. It is interesting that the conjecture holds good only for pure (for a fixed $m$) Lovelock but not for general (sum over $m$)  Einstein-Lovelock black holes. Like some other features \cite{Dadhich:2015lra,Dadhich:2012cv,Dadhich:2012pd}, the equipartition of gravitational and non-gravitational energy defining the black hole horizon is also yet another discriminator of pure Lovelock gravity. 
\section*{Acknowledgements}

Research of S.C. is funded by a SPM fellowship from CSIR, Government of India. A part of the work was done while ND was visiting  Albert Einstein Institute, Golm and University of KwaZulu-Natal, Durban, and for that he thanks respectively Hermann Nicolai and Sunil Maharaj.  
\bibliography{Gravity_1_full,Gravity_2_partial}

\bibliographystyle{./utphys1}
\end{document}